# Validity condition of the Jarzynski's relation for a classical mechanical system

Jaeyoung Sung[*]

*Department of Chemistry, Chung-Ang University, Seoul 156-756 Korea*

**Abstract**

Recently, Jarzynski suggested a striking thermodynamic equation that relates free energy change of a system and work done on the system during arbitrary nonequilibrium processes, which has been believed to hold irrespective of detailed nature of the nonequilibrium process. However, we show here that the Jarzynski's equation does not hold for an adiabatic process unless the phase-space extension of the system on completion of the adiabatic process coincides with that of the final equilibrium state of the system. This condition can be satisfied only when the adiabatic process does not change parameters on which the equilibrium phase-space extension of our system is dependent.

---

[*] email: jaeyoung@cau.ac.kr

Free energy, $F$, is one of the central concepts in thermodynamics, by which we can predict the direction of a spontaneous change between thermodynamic states of a system and the equilibrium subpopulation among various states of the system. However, it is often a difficult task to predict or to measure the free energy change quantitatively except for very simple systems, so quantification of the free energy of a complex system is one of the key issues in many problems of science. In conventional thermodynamics, it is well-established that the difference $\Delta F[\equiv F(\mathbf{B}) - F(\mathbf{A})]$ between the free energy of a system in thermodynamic state **A** and that of the system in thermodynamic state **B** is equal to the reversible work, $W_{rev}$ performed on the system during an infinitely slow isothermal change from state **A** to **B**. However, in practice, it is not easy to measure the reversible work directly from a finite time measurement.

Recently, Jarzynski put forward a striking relation of the free energy difference $\Delta F$ to statistical distribution $P(W)$ of the amount $W$ of work performed on the system during arbitrary nonequilibrium processes in the course of switching from state **A** to state **B**. The relation is given by

$$\langle \exp(-\beta W) \rangle = \exp(-\beta \Delta F), \tag{1}$$

where $\langle \ \rangle$ is defined by $\langle O \rangle \equiv \int dW O(W) P(W)$, and $\beta$ is $(k_B T)^{-1}$ with $k_B$ and $T$ being the Boltzmann constant and the absolute temperature, respectively.[1] Given that the system is initially in thermal equilibrium with a heat bath, Eq. (1) is argued to be valid irrespective of the shape of the path from **A** to **B** and the rate at which the change occurs along the path. Equation (1) was obtained first for a classical mechanical system undergoing an adiabatic process[1] and then for a system coupled to heat bath whose dynamics can be described by the classical or the quantum master equation.[2-5]

Until very recently Eq. (1) was believed to hold robustly irrespective of the detailed nature of the nonequilibrium process that changes the state of the system as is claimed in Ref. 1. Application of Eq. (1) to develop computational methods to calculate free energy change is recently discussed.[6,7] Furthermore, it is reported that Eq. (1) can be verified in the single RNA pulling experiment.[8]

However, very recently Cohen and Mauzerall claimed that the derivation of Eq. (1) is flawed for a system in contact with a heat bath,[9] Although the critiques in Ref. 9 was not made based on a rigorous proof, it made the validity of Eq. (1) a controversial issue.

In this Letter, we perform a theoretical analysis to check the validity of Eq. (1) for a classical mechanical system undergoing an adiabatic process, and obtain a validity condition of the Jarzynski's equation. It is found that, for a classical mechanical system suffering adiabatic processes, Eq. (1) holds if and only if the phase-space extension of the system on completion of the adiabatic process before a contact with a heat bath coincides with that of the final equilibrium state of the system, and that the latter condition can be satisfied only when the adiabatic process does not change any parameter on which the equilibrium phase-space extension of our system is dependent. For an example, as will be shown shortly, Eq. (1) does not hold for a gas system when the volume of the gas system is changed by an adiabatic process.

The free-energy of a system is a function of the parameters that we choose to identify a thermodynamic state of the system and the surrounding environment. These parameters may be called 'state parameters'. A thermodynamic state of a system defined by a set of state parameters can be represented by Gibbs ensemble of microscopic states of the system consistent with the set of state parameters. Let us consider the canonical ensemble of a classical gas system in which the state parameters are the temperature $T$ of a heat bath surrounding the system, the volume V in which the system is confined, and the number N of particles composing the system. N determines the dimensionality of the phase-space in which we can represent a microscopic state, say $\Gamma$, of the gas system as a single point. For a given value of $N$, the volume V determines the extension of the phase-space points representing the ensemble of microscopic states of our gas system. By definition, the gas system can assume only those microscopic states in the phase space extension, and dynamics of system occurs only in the phase-space extension. If we suddenly increase the volume of our gas system, for example, by expansion into vacuum, the equilibrium phase-space extension of the gas system after the vacuum expansion is greater than that before the expansion. It is well known that the free energy of the gas system after the expansion is lower than that before the expansion, which is the reason why the gas expansion into vacuum occurs spontaneously. However, during the process, the work done on the gas system is simply zero for every possible initial microscopic state in the initial phase space extension of the gas system. Therefore, we have $\langle \exp(-\beta W) \rangle = 1$ in contradiction with Eq. (1).[10] This example shows the fact that change in the free energy of a system can result from change in the volume of the phase-space extension of the system that can never be changed by performing a mechanical work on the classical system through an adiabatic process, which will be discussed in more detail shortly.

From now on, we examine Eq. (1) for more general adiabatic process of a classical mechanical system in which an arbitrary state parameter $V(t)$ of the system is varied with time $t$ in a controlled manner (with a constant rate, for example) from $V_0$ to $V_1$ for time $t_s$. We consider the same situation as that considered in Ref. 1, where the system was in thermal equilibrium with a heat bath before the system suffers an adiabatic process but is isolated from the heat bath during the adiabatic process. After the adiabatic process, the system is coupled again to the heat bath to get in thermal equilibrium with the heat bath without any mechanical work done on the system. Therefore, before and after the whole process, the temperature of our system is the same, for which case the free energy change $\Delta F$ during the whole process is well known as:

$$\exp[-\beta \Delta F] = \frac{\int_{\Omega_1} d\Gamma \exp[-\beta H_1(\Gamma)]}{\int_{\Omega_0} d\Gamma_0 \exp[-\beta H_0(\Gamma_0)]}, \qquad (2)$$

where $H_j(\Gamma)$ and $\int_{\Omega_j} d\Gamma$ respectively denote the Hamiltonian and the sum over all microscopic states or phase space extension of our system with state parameter $V_j$.

Now we obtain the expression for the L.H.S. of Eq. (1) for the same process. The work done on the system during the whole process is equal to the work done on the system during the adiabatic process that occurs for time $t_s$ as it does not cost any further work in the subsequent thermal process to get the system in thermal equilibrium with the heat bath. Let our expansion process begin at time 0 at which the microscopic state of our system happens to be $\Gamma_0$. $G(\Gamma, t | \Gamma_0) d\Gamma$ is to designate the conditional probability that we find our system between $\Gamma$ and $\Gamma + d\Gamma$ in phase space at time $t$, given that the system was at $\Gamma_0$ initially. In terms of $G(\Gamma, t | \Gamma_0)$, the average $\langle \exp(-\beta W) \rangle_{\Gamma_0}$ of $\exp(-\beta W)$ over $\Gamma$ for the system with the initial microscopic state being $\Gamma_0$ is written as

$$\langle \exp(-\beta W) \rangle_{\Gamma_0} = \int d\Gamma\, G(\Gamma, t_s | \Gamma_0) \exp(-\beta W(\Gamma | \Gamma_0)), \qquad (3)$$

where $W(\Gamma | \Gamma_0)$ denotes the amount of work done on the system during the adiabatic process in which the microscopic state of the system evolves from $\Gamma_0$ to $\Gamma$. Since the amount of work done on the system during an adiabatic process is equal to the change in the mechanical energy of the system, we have

$$W(\Gamma | \Gamma_0) = H_1(\Gamma) - H_0(\Gamma_0). \tag{4}$$

Furthermore, because the classical dynamics of an isolated system is deterministic, the phase space trajectory of our system during the adiabatic process is unique for a given initial microscopic state $\Gamma_0$, i.e.

$$G(\Gamma, t | \Gamma_0) = \delta(\Gamma - \Gamma^*(t | \Gamma_0)). \tag{5}$$

Here, $\delta(x)$ denotes the Dirac delta function, and $\Gamma^*(t | \Gamma_0)$ denotes the unique microscopic state of our system at time $t$ evolved from initial microscopic state $\Gamma_0$. By substituting Eq. (4) and (5) into Eq. (3), we have

$$\langle \exp(-\beta W) \rangle_{\Gamma_0} = \exp\{-\beta [H_1(\Gamma^*(t_s | \Gamma_0)) - H_0(\Gamma_0)]\}. \tag{6}$$

Since $\langle \exp(-\beta W) \rangle$ is equal to the average of $\langle \exp(-\beta W) \rangle_{\Gamma_0}$ over the initial equilibrium distribution of $\Gamma_0$, we have

$$\langle \exp(-\beta W) \rangle = \int_{\Omega_0} d\Gamma_0 \langle \exp(-\beta W) \rangle_{\Gamma_0} \frac{\exp(-\beta H_0(\Gamma_0))}{\int_{\Omega_0} d\Gamma'_0 \exp(-\beta H_0(\Gamma'_0))}. \tag{7}$$

Substituting Eq. (6) into Eq. (7), we get the expression for the L.H.S. of Eq. (1) as:

$$\langle \exp(-\beta W) \rangle = \frac{\int_{\Omega_0} d\Gamma_0 \exp(-\beta H_1(\Gamma^*(t_s | \Gamma_0)))}{\int_{\Omega_0} d\Gamma'_0 \exp(-\beta H_0(\Gamma'_0))}. \tag{8}$$

Comparing Eqs. (2) and (8), one can see that Eq. (1) holds if and only if

$$\frac{\int_{\Omega_0} d\Gamma_0 \exp(-\beta H_1(\Gamma^*(t_s | \Gamma_0)))}{\int_{\Omega_1} d\Gamma \exp(-\beta H_1(\Gamma))} = 1. \tag{9}$$

Because of the uniqueness of mechanical motion we can think of the initial phase point $\Gamma_0$ as a function of the phase-space point $\Gamma^*$ at time $t_s$, i.e. $\Gamma_0 = \Gamma_0(t_s, \Gamma^*)$ so we can change the integration variable from $\Gamma_0$ to $\Gamma^*$ as follows:

$$\int_{\Omega_0} d\Gamma_0 = \int_{\Omega(t_s)} d\Gamma^* \left|\frac{\partial \Gamma_0}{\partial \Gamma^*}\right|. \qquad (10)$$

Here, $\left|\frac{\partial \Gamma_0}{\partial \Gamma^*}\right|$ denotes the Jacobian determinant and $\int_{\Omega(t_s)} d\Gamma^*$ denotes sum over the phase space extension of our system at time $t_s$ at the very end of the adiabatic process before the thermal process. According to Gibbs,[11] the volume of the phase-space extension available to a mechanical system does not change in time although shape of the boundary of the phase-space extension will change according to the dynamics of system, which is known as the principle of conservation of extension in phase.[12] The principle is true as far as dynamics of the system obey the Hamilton's equation of motion or the Liouville equation so that it holds even for a system with a time dependent potential function. Therefore, we have the following identity

$$\int_{\Omega_0} d\Gamma_0 = \int_{\Omega(t_s)} d\Gamma^* \qquad (11)$$

or equivalently,

$$\left|\frac{\partial \Gamma_0(t_s)}{\partial \Gamma^*}\right| = 1. \qquad (12)$$

With this at hand, we can rewrite the validity condition of Eq. (1), given in Eq. (9) as

$$\frac{\int_{\Omega(t_s)} d\Gamma \exp(-\beta H_1(\Gamma))}{\int_{\Omega_1} d\Gamma \exp(-\beta H_1(\Gamma))} = 1. \qquad (13)$$

Remember that, whereas $\int_{\Omega(t_s)} d\Gamma$ denotes the sum over the phase space extension of our system with $V = V_1$ at the very end of the adiabatic process before the subsequent thermal process to get the system in thermal equilibrium, $\int_{\Omega_1} d\Gamma$ denotes that of the system with $V = V_1$ in equilibrium with the heat bath. Eq. (13) indicates that the Jarzynski's relation given by Eq. (1) holds if and only if the phase space extension of the system with $V = V_1$ at thermal equilibrium coincides with that of the system at time $t_s$ at which the adiabatic process to change the state parameter from $V_0$ to $V_1$ is just completed.

Noting the principle of extension in phase given in Eq. (11), one can see that the validity condition, Eq. (13), of Jarzynski's relation cannot be satisfied unless

$$\frac{\int_{\Omega_0} d\Gamma_0}{\int_{\Omega_1} d\Gamma} = 1. \qquad (14)$$

Equation (14) is a necessary condition for the Jarzynski equation to hold. Note that Eq. (14) can be obtained also directly from the high temperature limit of Eq. (9).

Equation (14) tells us that free energy difference between states with different phase space extension of a system cannot be estimated by Jarzynski relation or by Eq. (1). For an example, let's consider free energy difference $\Delta F_{id}(\xi)$ of a system of one-dimensional ideal gas particles with speed less than $\xi$ confined in a box with length $L_1$ from the same system confined in the box with length $L_0$. For the system composed of $N$ ideal gas particles with a unit mass, the L.H.S. of Eq. (14) is given by

$\dfrac{\prod_{j=1}^{N} \int_{-\xi}^{\xi} dp_j \int_{0}^{L_1} dx_j}{\prod_{j=1}^{N} \int_{-\xi}^{\xi} dp_j \int_{0}^{L_0} dx_j} = \left(\dfrac{L_1}{L_0}\right)^N$ ; therefore, the necessary condition given in Eq. (14) does not hold unless $L_0$ is equal to $L_1$ so that Jarzynski relation given in Eq. (1) does not hold, neither.

For the ideal gas model discussed in the previous paragraph, direct comparison between L.H.S. and R.H.S. of Eq. (1) can be made, which indicates that Eq. (1) breaks down. The L.H.S. of Eq. (1), $\langle \exp(-\beta W) \rangle$ is dependent on $\xi$. For incidence, for the adiabatic ideal gas expansion process in which position of the right wall of the gas system increases from $L_0$ to $L_1$ during time $t_S$ in constant rate $v_P$ with the left wall held fixed at the coordinate origin, the explicit expression for $\langle \exp(-\beta W) \rangle$ is obtained as follows:

$$\langle \exp(-\beta W) \rangle = (L_1/L_0)^N \left\{ \frac{1 - R\,\mathrm{Erfc}\left(\sqrt{\beta m [\xi - 2M(\xi)v_P]^2/2}\right) + \chi(\xi)}{\mathrm{Erf}\left(\sqrt{\beta m \xi^2/2}\right)} \right\}^N . \quad (15)$$

where $\chi(\xi) = \dfrac{1}{2\sqrt{a\pi}} \left\{ \phi[\sqrt{a}(2M(\xi)R + 1 - R)] - \phi[\sqrt{a}(2M(\xi)R - 1 + R)] \right\}$ with $a = \beta(L_1/t_S)^2/2$, $R = L_0/L_1$, and $\phi(z) = \exp(-z^2) - z\,\mathrm{Erfc}(z)$. Here, $M(\xi)$ denotes the number of collisions made by the ideal gas particle with initial speed $\xi$ to the moving wall during time interval $(0, t_S)$; the value of $M$ is given by the greatest integer that is not greater than $2^{-1}[(\xi t_S + x_0)/L_1 + 1]$ with $|x_0| < L_0$,[13] where $x_0$ is the initial position of the gas particle. In comparison, $\exp(-\beta \Delta F_{id}(\xi))$ is given by $\exp(-\beta \Delta F_{id}(\xi)) = (L_1/L_0)^N$ irrespective of $\xi$ according to conventional statistical mechanics. Therefore, the Jarzynski relation in Eq. (1) does not hold for this system unless $L_1$ is equal to $L_0$ or $t_S = 0$, in which case both $\langle \exp(-\beta W) \rangle$ given in Eq. (15) and $\exp(-\beta \Delta F_{id})$ become unity.

It should be mentioned that the necessary condition given in Eq. (14) is strictly applicable only to the case where the volume $\int_{\Omega_0} d\Gamma_0$ of the initial phase-space extension is finite. When $\int_{\Omega_0} d\Gamma_0$ is infinite, even if Eq. (14) is not satisfied, $\int_{\Omega(t_S)} d\Gamma$ can coincide with $\int_{\Omega_1} d\Gamma_1$ and the necessary and the sufficient condition given in Eq. (13) of Jarzynski relation can be satisfied without breaking the principle of conservation in phase. For incidence, for our ideal gas system discussed in the previous paragraph, $\langle \exp(-\beta W) \rangle$ given in Eq. (15) reduces to $\exp(-\beta \Delta F_{id})$ in the infinite $\xi$ limit, in accordance with recent work of Lua and Grosberg for a one-dimensional ideal gas system.[14] For the case with infinite $\xi$, every initial phase points in $\int_{\Omega_0} d\Gamma_0$ can be connected one by one to every phase points in $\int_{\Omega_1} d\Gamma_1$ at any time $t_S$ by a phase point trajectory during the adiabatic expansion occurring in a constant rate so that the condition given in Eq. (13) is satisfied,[15] except for the case with the vacuum expansion in which case one can show that Eq. (13) no longer holds.

In practice, it is impossible for us to sample the entire range of an infinite phase-space extension of a system so that the range of phase space extension we sample initially is always finite in any experiment or computer simulation, however large the size of the sampling is. Therefore, the experimental estimation of the free energy difference based on Jarzynski relation can never be exact even in the absence of any experimental error. For our ideal gas model considered above, as the initial velocity distribution of a gas particle follows Gaussian, there always exists a microscopic state with the maximum speed $v_{max}$ among sampled microscopic states in an experiment; equivalently, there always exist microscopic states with speed greater than $v_{max}$ whose contribution to experimental estimation of the L.H.S. of Eq. (1) is neglected.

In the present Letter, we present a theoretical test of the Jarzynski's relation of free energy change of a system to the work done on the system during an adiabatic process for which dynamics of the system obeys classical mechanics. For the system, we find the validity condition, Eqs. (13) and (14) of the Jarzynski equation. This condition can be satisfied only when the adiabatic process does not change any parameter on which the equilibrium phase-space extension of our system is dependent. Research to find the validity condition of the Jarzynski's relation for a system with a stochastic dynamics is also under progress.

**ACKNOWLEDGEMENT**

The author would like to acknowledge R. J. Silbey, E. G. D. Cohen, and C. Jarzynski gratefully for helpful comments. This research is supported by Chung-Ang University Research grants in 2004.